\documentclass[lettersize,journal]{IEEEtran}
\usepackage[final]{graphicx}
\usepackage{subfigure}
\usepackage{float}
\usepackage{amsmath}
\usepackage{cases}
\usepackage{color}
\usepackage{colortbl}
\usepackage{algorithm}
\usepackage{algpseudocode}
\usepackage{amsmath}
\usepackage{amssymb}
\usepackage{booktabs}
\usepackage{setspace}
\usepackage{array}
\usepackage{rotating}
\usepackage[normalem]{ulem}
\usepackage{bbding}

\renewcommand{\normalsize}{\fontsize{10}{12}\selectfont}

\makeatletter
\renewcommand{\maketag@@@}[1]{\hbox{\m@th\normalsize\normalfont#1}}%
\makeatother
\usepackage{stfloats}
\usepackage{cite}
\usepackage{makecell}
\usepackage{multirow}
\usepackage{hyperref}

\ifCLASSINFOpdf
\else
\fi

\usepackage{caption}
\usepackage{subcaption}
\usepackage{mathtools}
\usepackage{etoolbox}  

\begin{document}
\title{DMSNet: Cross-Band Learning for Multi-Target Sensing in Multi-Band ISAC}
\author{Haotian Liu,~\IEEEmembership{Graduate Student Member,~IEEE,}
Zhiqing Wei,~\IEEEmembership{Member,~IEEE,}
Quanjiang Zhao,
Lin Wang,~\IEEEmembership{Graduate Student Member,~IEEE,}
Yunxin Geng,~\IEEEmembership{Graduate Student Member,~IEEE,} \\
Xingwang Li,~\IEEEmembership{Senior Member,~IEEE,}
Zhiyong Feng,~\IEEEmembership{Senior Member,~IEEE}

\thanks{Haotian Liu, Zhiqing Wei, Quanjiang Zhao, Lin Wang, Yunxin Geng, and Zhiyong Feng are with the Key Laboratory of Universal Wireless Communication, Ministry of Education, Beijing University of Posts and Telecommunications, Beijing 100876, China (emails: \{haotian\_liu; weizhiqing; qjzhao; wlwl; gengyx; fengzy\}@bupt.edu.cn). \textit{Corresponding authors: Zhiqing Wei, Zhiyong Feng.}

Xingwang Li is with the School of Physics and Electronic Information Engineering, Henan Polytechnic University, Jiaozuo 454000, China (e-mail: lixingwangbupt@gmail.com).}}

\maketitle

\begin{abstract}
Multi-band integrated sensing and communication (ISAC) offers complementary high- and low-frequency echo information for multi-target sensing.
However, existing dual-band ISAC sensing methods have a limited ability to exploit deep complementary information across heterogeneous bands and often incur high computational costs.
To address these limitations, we propose a Dual-Band Multi-Target Sensing Neural Network (DMSNet) for joint target number and parameter estimation.
Under representative simulation conditions, DMSNet outperforms the best baseline in target number estimation, increasing count accuracy from 89.01\% to 91.74\% and Macro-F1 from 90.80\% to 93.07\%.
For parameter estimation, compared with the best baselines, DMSNet reduces the median absolute errors of range, velocity, and angle by 82.2\%, 56.9\%, and 73.2\%, respectively.
Moreover, DMSNet reduces runtime by 68.7\% relative to the fastest existing dual-band ISAC sensing method.
\end{abstract}

\section{Introduction}
% 面向低空安防、智能交通等新兴场景，具备 ISAC 能力的无线网络需要提供可靠的多目标感知能力【Fan Liu】。然而，单一频段受其频谱资源和频率相关物理特性的制约，难以充分适应复杂感知环境[IOTM, ICCC]。高频与低频在分辨能力、传播特性及目标散射响应等方面呈现显著差异，并具有互补潜力【ICCC】。因此，多频段 ISAC 中的高低频协作感知受到广泛关注【四篇CA工作】。
Emerging scenarios such as low-altitude economy and intelligent transportation require ISAC-enabled wireless networks to provide reliable multi-target sensing capabilities~\cite{Liu2022}. 
However, single-band ISAC is constrained by limited spectrum resources and the frequency-dependent characteristics of sensing targets and propagation environments, making it difficult to fully adapt to complex sensing scenarios~\cite{Limagazine}.
High- and low-frequency bands exhibit significant differences in resolution capability, propagation characteristics, and target scattering responses, and thus offer complementary sensing potential~\cite{Limagazine}. 
Therefore, high- and low-frequency cooperative sensing in multi-band ISAC has attracted increasing attention~\cite{Wei2024CA,zhangCA,Liu2025CA,CaiCA}.

% 面向多目标感知，高低频协作 ISAC 需同时实现目标数量与参数估计。2024 年，Wei 等率先研究单目标 OFDM 场景，通过融合高低频感知谱并结合压缩感知处理实现距离和速度估计~\cite{Wei2024CA}。同年，Zhang 等进一步面向具有异构参数配置的多频 OFDM 系统，设计联合多频处理方法，实现跨频距离和速度估计~\cite{zhangCA}。2025 年，Liu 等将研究扩展至多目标 MIMO-OFDM 场景，提出数据级与符号级融合方法，实现角度、距离和速度联合估计~\cite{Liu2025CA}。近期，Cai 等进一步考虑城市低空杂波环境，通过三阶张量分解提取多维目标特征，并结合跨频因子融合实现目标参数估计~\cite{CaiCA}。然而，现有方法多依赖预定义的融合机制，难以充分挖掘异构高低频观测中的深层互补信息，且计算开销较高，多数方法亦未考虑目标数量估计。
For multi-target sensing, the sensing task involves both target number estimation and target parameter estimation.
In 2024, Wei \textit{et al.} first investigated a single-target OFDM scenario, where traditional fast Fourier transform (FFT) and compressive sensing were used to fuse high- and low-frequency sensing echoes and estimate the range and velocity of the target~\cite{Wei2024CA}. 
Furthermore, Zhang \textit{et al.} proposed an improved FFT method to achieve the fusion of multi-band sensing information as well as target parameter estimation~\cite{zhangCA}. 
In 2025, Liu \textit{et al.} extended the study to multi-target MIMO-OFDM scenarios and proposed data-level and symbol-level fusion methods for joint angle, range, and velocity estimation~\cite{Liu2025CA}. 
More recently, Cai \textit{et al.} further extracted multidimensional target features through CANDECOMP/PARAFAC (CP) decomposition and performed cross-band factor fusion for target parameter estimation~\cite{CaiCA}. 
However, existing methods largely rely on handcrafted fusion rules, making it difficult to exploit deep complementary information from heterogeneous high- and low-frequency signals.
Moreover, iterative reconstruction, matrix decomposition, and extensive spectral search often lead to high computational costs, while most methods~\cite{Wei2024CA,zhangCA,CaiCA} do not consider target number estimation.

% 近年来，基于AI的ISAC 感知逐渐用于复杂多目标任务。代表性地，CSIYOLO 通过神经网络学习多尺度目标特征，实现多目标检测与参数估计~\cite{zhang2025csiyolo}。然而，现有学习式方法尚未考虑异构高低频观测的协同融合。因此，如何自适应融合高低频信息并联合实现目标数量与参数估计，仍有待研究。
Recently, learning-based ISAC sensing has been increasingly applied to complex multi-target sensing tasks. 
As a representative approach, CSIYOLO learns multi-scale target features through neural networks to achieve multi-target number and parameter estimation~\cite{zhang2025csiyolo}. 
However, existing learning-based methods have not considered the multi-band ISAC scenarios. 
Therefore, how to deeply exploit and adaptively fuse complementary high- and low-frequency sensing information for joint target number and parameter estimation remains an open problem.

% 针对上述问题，本文首次提出一种面向异构高低频协作 ISAC 的双频多目标感知神经网络（DMSNet），联合实现目标数量以及距离、速度和角度参数估计。该网络通过跨频特征交互自适应挖掘高低频观测的互补信息，并采用目标数量引导的粗到细估计实现连续多目标参数恢复。仿真结果表明，DMSNet 在目标数量与参数估计方面均取得更优性能，同时显著降低了相较传统双频信号处理方法的运行时间。
To this end, we propose, for the first time, a Dual-Band Multi-Target Sensing Neural Network (DMSNet) for cooperative ISAC over heterogeneous high- and low-frequency bands, which jointly estimates the target number as well as range, velocity, and angle parameters.
DMSNet first estimates the number of targets through the target number estimation module, whose output guides the subsequent parameter estimation process.
The designed coarse and fine parameter estimation modules then progressively refine the range, velocity, and angle estimates for multiple targets. 
In all three modules, high- and low-frequency features are adaptively fused to effectively exploit the complementary sensing information across heterogeneous frequency bands.
Simulation results show that, at a representative SNR of $0$ dB, DMSNet improves Count Accuracy from 89.01\% to 91.74\% and Macro-F1 from 90.80\% to 93.07\%, while reducing MAE by 25.8\%.
Compared with the best baselines, it further reduces the median absolute errors of range, velocity, and angle estimation by 82.2\%, 56.9\%, and 73.2\%, respectively, while reducing runtime by 68.7\% relative to the fastest conventional dual-band pipeline~\cite{CaiCA}.

\textit{Notations:}
Vectors and matrices are written in bold letters and in capital bold letters, respectively. 
$\mathbb{C}$ and $\mathbb{R}$ denote the set of complex and real numbers.
$|\cdot|$ and $\odot$ denote the absolute operator and Hadamard product, respectively.

\section{System Model}\label{se2}
We consider a multi-band ISAC base station (BS) for multi-target sensing.
The multi-band ISAC signal aggregates $B$ non-contiguous frequency bands for communication and sensing based on carrier aggregation technology~\cite{Wei2024CA,Liu2025CA}. 
In the sensing stage, each band $b\in\mathcal{B}=\{1,2,\cdots,B\}$ occupies $N_\mathrm{c}^b$ subcarriers and $M_\mathrm{s}^b$ known OFDM pilot symbols.
Based on the 3GPP standard, each non-contiguous frequency band employs an independent RF chain and a per-band uniform linear array (ULA) antenna~\cite{3GPP38300}, where the numbers of transmit and receive antennas are $N_\mathrm{t}^b$ and $N_\mathrm{r}^b$. 
We assume that there are at most $L_\mathrm{max}$ targets in the region of interest, and the actual number of targets is $L\le L_\mathrm{max}$.

\subsection{Multi-band ISAC Sensing Echo Model}
The $l\in\{1,2,\cdots,L\}$-th target is described by its range $r_l$, radial velocity $v_l$, and angle $\theta_l$.
For the $b$-th band, the received echo vector at the $n$-th subcarrier and the $m$-th OFDM symbol can be expressed as~\cite{Liu2025CA}
\begin{equation}\label{eq1}
\begin{aligned}
  \mathbf{y}_{n,m}^b = &\sqrt{P_b}s_{n,m}^b\sum_{l=1}^L
  \sqrt{\frac{\lambda_b^2}{(4\pi)^3r_l^4}}\rho_{b,l}e^{j\phi_{b,l}}\chi_{b,l}\mathbf{a}_b\left(\theta_l\right) \\ &
  \times e^{-j2\pi n \Delta f_b \tau_l}e^{j2\pi m T_b f_{\mathrm{d},l}^b} + \mathbf{z}_{n,m}^b,
\end{aligned}
\end{equation}
where $P_b$ is the transmit power, $s_{n,m}^b$ is the known pilot symbol, and $\mathbf{z}_{n,m}^b\in\mathbb{C}^{N_\mathrm{r}^b\times 1}$ denotes the additive white Gaussian noise (AWGN);
$\lambda_b=\frac{c_0}{f_\mathrm{c}^b}$ is the wavelength, where $c_0$ and $f_\mathrm{c}^b$ represent the speed of light and carrier frequency, respectively;
$\rho_{b,l}$ is the frequency-selective complex scattering coefficient, and the corresponding radar cross section (RCS) is denoted by $\xi_{b,l}=|\rho_{b,l}|^2$~\cite{zhangRCS-2026}.
This indicates that a target weakly visible in one band may still be detectable in another band, which provides a key motivation for multi-band target detection~\cite{zhangRCS-2026};
$\phi_{b,l}$ is the band-dependent phase offset, and $\chi_{b,l}$ denotes the beamforming gain;
$\tau_l=2r_l/c_0$ and $f_{\mathrm{d},l}^b=2f_\mathrm{c}^bv_l/c_0$ denote the delay and Doppler shift, respectively;
$\Delta f_b$ and $T_b$ represent the subcarrier spacing and total symbol duration, respectively;
$\mathbf{a}_b\left(\cdot\right)$ is the receive steering vector, expressed as
\begin{equation}\label{eq2}
 \mathbf{a}_b\left(\cdot\right)=\left[1,e^{j2\pi \frac{d_b}{\lambda_b}\sin(\cdot)},\cdots,e^{j2\left(N_\mathrm{r}^b-1\right)\pi \frac{d_b}{\lambda_b}\sin(\cdot)}\right]^\mathrm{T},   
\end{equation}

After pilot removal and power normalization, all elements across subcarriers, symbols, and antennas are stacked into a tensor $\mathcal{X}_b\in\mathbb{C}^{N_\mathrm{c}^b\times M_\mathrm{s}^b \times N_\mathrm{r}^b}$, and its $(n,m,k)$-th element is expressed as
\begin{equation}\label{eq3}
\begin{aligned}
\left[\mathcal{X}_b\right]_{n,m,k}=&\sum_{l=1}^L\sqrt{\frac{\lambda_b^2}{(4\pi)^3r_l^4}}\rho_{b,l}e^{j\phi_{b,l}}\chi_{b,l}e^{-j2\pi n \Delta f_b \tau_l}\\ & \times e^{j2\pi m T_b f_{\mathrm{d},l}^b}e^{j2\pi k \frac{d_b}{\lambda_b}\sin(\theta_l)} + \tilde{z}_{n,m,k}^b      
\end{aligned} 
\end{equation}

High- and low-frequency cooperation is a representative yet challenging multi-band ISAC case, where high-frequency bands provide finer resolution and low-frequency bands offer more reliable observations~\cite{Wei2024CA,Liu2025CA,zhangCA,CaiCA}. 
However, their heterogeneous signal structures, array responses, Doppler scales, phase offsets, and scattering responses hinder deep fusion. 
Therefore, we design a neural network to exploit cross-band complementary features for joint multi-target number and parameter estimation from $\left\{\mathcal{X}_b\right\}_{b\in\{\mathrm{h},\mathrm{l}\}}$.

\begin{figure}
    \centering
    \includegraphics[width=0.5\textwidth]{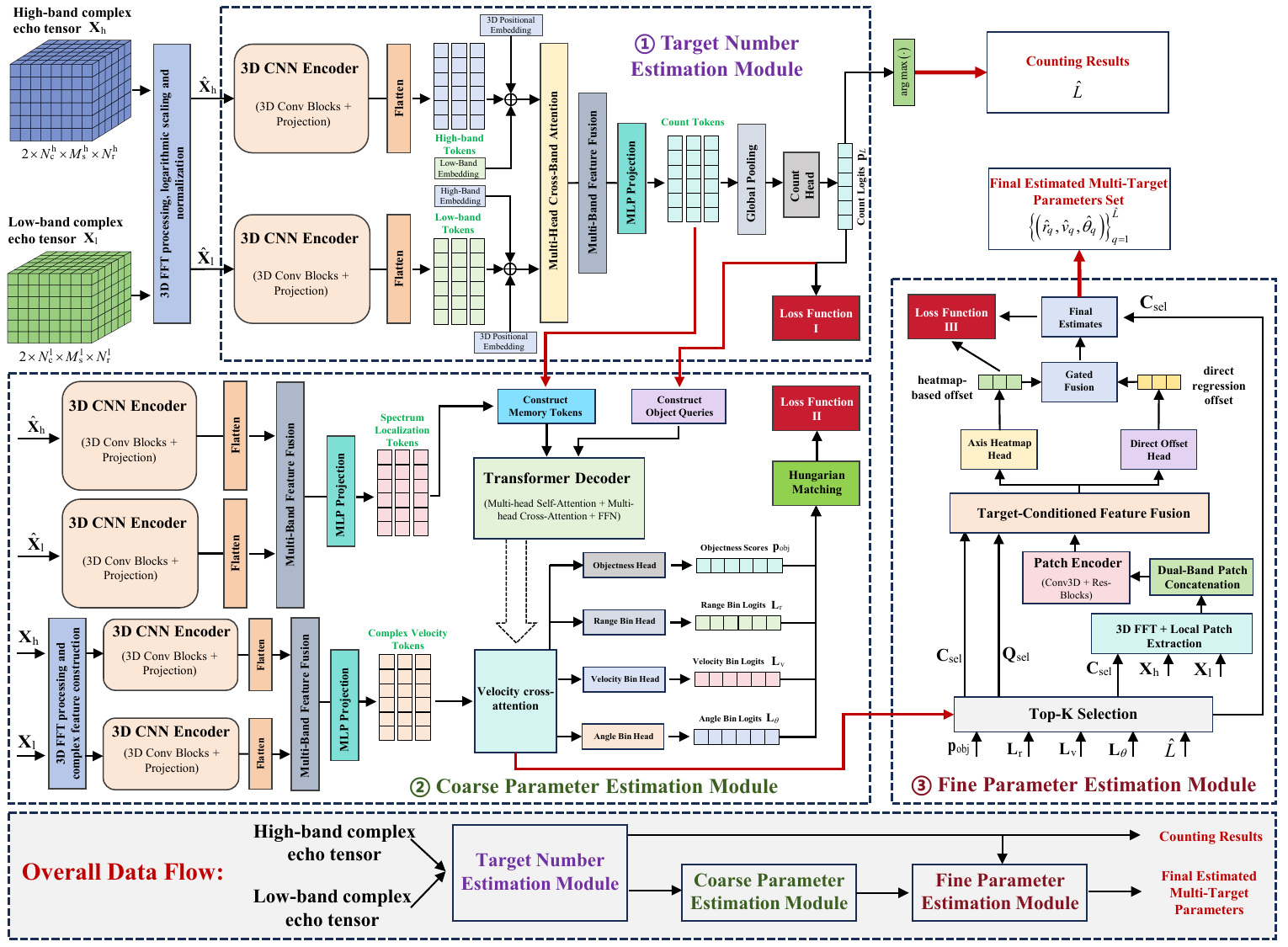}
    \caption{The illustration of the proposed DMSNet.}
    \label{fig:dmsnet}
\end{figure}

\section{Proposed Dual-Band Multi-Target Sensing Neural Network (DMSNet)}\label{se3}

As shown in Fig.~\ref{fig:dmsnet}, DMSNet consists of three cascaded modules for target number estimation, coarse parameter estimation, and fine parameter estimation.

\subsection{Overall Architecture}\label{se3-a}

DMSNet follows a coarse-to-fine inference pipeline. 
Its inputs are the high- and low-band complex echo tensors
$\{\mathbf{X}_{b}\in\mathbb{R}^{2\times N_\mathrm{c}^b\times M_\mathrm{s}^b\times N_\mathrm{r}^b}\}_{b\in\{\mathrm{h},\mathrm{l}\}}$,
where the first dimension contains the real and imaginary channels.
The target number module first estimates
$\hat L\in\{0,\ldots,L_{\max}\}$.
The coarse module then converts multi-target parameter estimation into classification over discretized parameter grids, producing coarse target candidates.
Around these candidates, the fine module extracts local dual-band spectral patches and estimates residual offsets through local feature regression.
Finally, DMSNet outputs
$\{(\hat r_p,\hat v_p,\hat\theta_p)\}_{p=1}^{\hat L}$.
Throughout DMSNet, the two bands are encoded separately and fused at the feature level to preserve band-specific information.

\subsection{Target Number Estimation Module}\label{se3-b}

Each complex echo tensor $\mathbf{X}_b$ is converted into a normalized range-velocity-angle power spectrum $\hat{\mathbf{X}}_b$ through three-dimensional (3D) FFT processing, logarithmic scaling, and normalization, followed by a 3D CNN encoder.

Different from conventional multi-band fusion using hand-crafted or SNR-based weights~\cite{Wei2024CA,Liu2025CA,zhangCA,CaiCA}, this module adaptively learns cross-band complementarity at the feature level.
The encoded features are flattened into high- and low-band tokens
$\{\mathbf{T}_b\in\mathbb{R}^{512\times128}\}_{b\in\{\mathrm{h},\mathrm{l}\}}$,
and multi-head cross-band attention~\cite{vaswani2017attention} is expressed as
\begin{equation}\label{eq4}
    \bar{\mathbf{T}}_{b}
    =
    \mathbf{T}_{b}
    +
    {\rm MHA}(\mathbf{T}_{b},\mathbf{T}_{\bar b},\mathbf{T}_{\bar b}),
    \quad b\in\{\mathrm{h},\mathrm{l}\},
\end{equation}
where $\bar b$ denotes the other band.
The enhanced tokens are further fused as
\begin{equation}\label{eq5}
    \mathbf{U}_{\rm num}
    =
    \left[
    \bar{\mathbf{T}}_\mathrm{h},
    \bar{\mathbf{T}}_\mathrm{l},
    |\bar{\mathbf{T}}_\mathrm{h}-\bar{\mathbf{T}}_\mathrm{l}|,
    \bar{\mathbf{T}}_\mathrm{h}\odot\bar{\mathbf{T}}_\mathrm{l}
    \right],
\end{equation}
where $|\cdot|$ and $\odot$ denote element-wise absolute value and the Hadamard product, respectively; the four terms preserve single-band features and characterize cross-band discrepancy and consistency.
The count tokens are obtained as
$\mathbf{T}_{\rm num}={\rm MLP}(\mathbf{U}_{\rm num})\in\mathbb{R}^{512\times128}$.
After pooling and a count head, the module outputs
$\mathbf{p}_{L}\in\mathbb{R}^{L_{\max}+1}$ and estimates
$\hat L=\underset{l}{\arg\max}[\mathbf{p}_{L}]_l$.

\subsection{Coarse Parameter Estimation Module}\label{se3-c}

The coarse module uniformly divides each parameter domain
$[d_{\min},d_{\max}]$, $d\in\{r,v,\theta\}$, into $G_d$ bins and predicts a fixed number of candidate targets.
This classification-based design avoids direct continuous regression, alleviates target-order ambiguity, and provides reliable candidates for fine refinement.

As shown in Fig.~\ref{fig:dmsnet}, three complementary token types are exploited:
the count tokens $\mathbf{T}_{\rm num}$ provide target-number context,
the spectrum localization tokens
$\mathbf{T}_{\rm spec}\in\mathbb{R}^{512\times192}$
characterize coarse range-velocity-angle peak structures,
and the complex velocity tokens
$\mathbf{T}_{\rm vel}\in\mathbb{R}^{512\times192}$
retain Doppler-related phase cues weakened in power-spectrum features.

After dimension alignment, $\mathbf{T}_{\rm num}$ and $\mathbf{T}_{\rm spec}$ construct the memory tokens, while $\mathbf{p}_{L}$ is used to generate object queries.
A Transformer decoder extracts target-wise features from the memory tokens and object queries, which further interact with $\mathbf{T}_{\rm vel}$ through velocity cross-attention.
The resulting query features
$\mathbf{Z}_{\rm coarse}\in\mathbb{R}^{L_{\max}\times192}$
are fed into four prediction heads:
\begin{equation}\label{eq6}
    \left\{
    \mathbf{p}_{\rm obj}(p),
    \mathbf{L}_{r}(p),
    \mathbf{L}_{v}(p),
    \mathbf{L}_{\theta}(p)
    \right\}_{p=1}^{L_{\max}},
\end{equation}
where $\mathbf{p}_{\rm obj}(p)\in[0,1]$ is the objectness score and
$\mathbf{L}_{d}(p)\in\mathbb{R}^{G_d}$ is the bin-logit vector for
$d\in\{r,v,\theta\}$.
The query features
obtained by the velocity cross-attention module are retained for fine refinement.

\subsection{Fine Parameter Estimation Module}\label{se3-d}

According to $\hat L$, the top-$\hat L$ candidates are selected by their objectness scores, and their predicted bins are converted into physical coarse centers
$\{\mathbf{c}_p\}_{p=1}^{\hat L}$.

Instead of directly regressing continuous parameters from the full echo tensors, the fine module performs coarse-guided local refinement.
The high- and low-band echoes are independently transformed by $256\times256\times256$ 3D FFT, and local spectral patches are extracted around each coarse center using band-specific physical mappings.
The dual-band patches are then concatenated to exploit complementary local magnitude and phase structures.

The concatenated patches are encoded by a patch encoder and further fused with the selected query features $\mathbf{Q}_{\rm sel}$ and coarse-center information $\mathbf{C}_{\rm sel}$ through target-conditioned feature fusion.
An axis heatmap head and a direct offset head estimate axis-wise peak locations and coupled 3D residual information, respectively.
Their outputs are adaptively fused as
\begin{equation}\label{eq7}
    \hat{\delta}_{d,p}
    =
    (1-g_{d,p})\delta^{\rm hm}_{d,p}
    +
    g_{d,p}\delta^{\rm dir}_{d,p},
    \quad d\in\{r,v,\theta\},
\end{equation}
where $\delta^{\rm hm}_{d,p}$ and $\delta^{\rm dir}_{d,p}$ denote the heatmap-based and direct regression offsets, respectively, and
$g_{d,p}\in[0,1]$ is the learned fusion gate.
The final estimate is
\begin{equation}\label{eq8}
    \hat{s}_{d,p}
    =
    c_{d,p}
    +
    \Delta_d\hat{\delta}_{d,p},
    \quad d\in\{r,v,\theta\},
\end{equation}
where
$\Delta_d=(d_{\max}-d_{\min})/G_d$
is the grid interval along dimension $d$.

\subsection{Training Objective}\label{se3-e}

Since the discrete target-number decision, top-$\hat L$ selection, and coarse-bin indexing interrupt cross-module gradient propagation, DMSNet is trained module-wise and cascaded during inference. 

For target number estimation, let
$\boldsymbol{\pi}_{L}={\rm softmax}(\mathbf{p}_{L})$.
The count loss is
\begin{equation}\label{eq9}
    \mathcal{L}_{\rm num}
    =
    -\sum_{k=0}^{L_{\max}}
    y_k^{L}\log \pi_{L,k},
\end{equation}
where $y_k^{L}=\mathbb{I}(L=k)$ is the one-hot target-number label and $\mathbb{I}(\cdot)$ denotes the indicator function.

For coarse estimation, Hungarian matching~\cite{Carion2020DETR} associates the unordered ground-truth targets with candidate queries.
Let $\mathcal{M}$ denote the set of matched query indices,
$y_p^{\rm obj}\in\{0,1\}$ the objectness label of query $p$,
and $b_{d,p}^{\rm gt}$ the ground-truth bin index of its matched target along dimension $d\in\{r,v,\theta\}$.
The coarse loss is
\begin{equation}\label{eq10}
{\fontsize{8}{8}\begin{aligned}
    \mathcal{L}_{\rm coarse}
    ={}&
    -\lambda_{\rm obj}
    \sum_{p=1}^{L_{\max}}
    \Big[
    y_p^{\rm obj}\log \mathbf{p}_{\rm obj}(p)
    +(1-y_p^{\rm obj})
    \log\!\big(1-\mathbf{p}_{\rm obj}(p)\big)
    \Big] \\
    &-\lambda_{\rm bin}
    \sum_{p\in\mathcal{M}}
    \sum_{d\in\{r,v,\theta\}}
    \log
    \left[
    {\rm softmax}\!\left(\mathbf{L}_{d}(p)\right)
    \right]_{b_{d,p}^{\rm gt}} .
\end{aligned}}
\end{equation}

For fine estimation, the normalized ground-truth offset is
$\delta_{d,p}^{\rm gt}=(s_{d,p}^{\rm gt}-c_{d,p})/\Delta_d$.
Let
$\boldsymbol{\pi}_{d,p}^{\rm hm}={\rm softmax}(\mathbf{h}_{d,p})$
denote the predicted axis-wise probability distribution and
$\mathbf{y}_{d,p}^{\rm hm}$
its Gaussian soft label centered at $\delta_{d,p}^{\rm gt}$.
The fine loss combines heatmap supervision and Smooth-$L_1$ offset regression~\cite{Girshick2015FastRCNN}:
\begin{equation}\label{eq11}
\begin{aligned}
    \mathcal{L}_{\rm fine}
    ={}&
    -\lambda_{\rm hm}
    \sum_{p\in\mathcal{M}}
    \sum_{d\in\{r,v,\theta\}}
    \sum_j
    y_{d,p}^{\rm hm}(j)
    \log \pi_{d,p}^{\rm hm}(j) \\
    &+\lambda_{\rm off}
    \sum_{p\in\mathcal{M}}
    \sum_{d\in\{r,v,\theta\}}
    {\rm SmoothL1}
    \left(
    \hat{\delta}_{d,p},
    \delta_{d,p}^{\rm gt}
    \right).
\end{aligned}
\end{equation}

\section{Simulation Results and Analysis}\label{se4}

\begin{figure}
    \centering
    \includegraphics[width=0.45\textwidth]{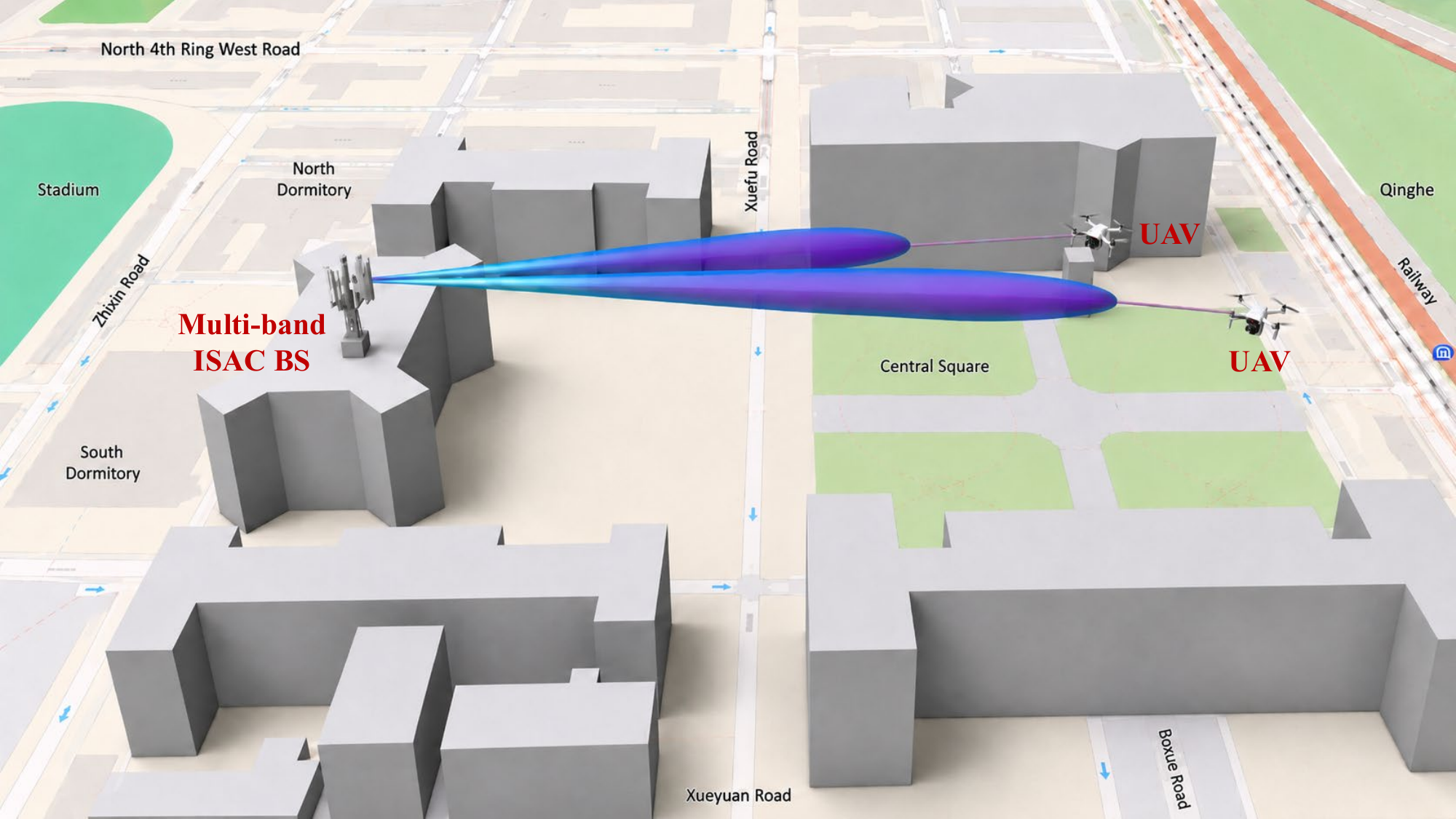}
    \caption{The 3D digital twin of the Beijing University of Posts and Telecommunications (BUPT) campus.}
    \label{fig:scenario}
\end{figure}
\subsection{Simulation Setup}\label{se4-a}

As shown in Fig.~\ref{fig:scenario}, a dual-band UAV sensing dataset is generated in a 1:1 3D digital twin of the Beijing University of Posts and Telecommunications (BUPT) campus using Sionna RT~\cite{Hoydis2023SionnaRT}.

The low- and high-band carrier frequencies are 3.5 and 28 GHz, respectively, with $N_\mathrm{c}^{b}=64$. 
Their $(\Delta f^{b},M_\mathrm{s}^{b},N_\mathrm{r}^{b})$ are $(30~\mathrm{kHz},14,4)$ and $(240~\mathrm{kHz},112,32)$, respectively~\cite{3GPP38300,Liu2025CA}.
Each scene contains $L\in\{0,\ldots,5\}$ UAVs with
$r\in[50,300]$ m, $v\in[-30,30]$ m/s, and
$\theta\in[-\pi/3,\pi/3]$.
Training SNRs range from $-10$ to $20$ dB in 5-dB steps.

The dataset contains $50,000$ samples with an 80\%/20\% training-validation split and an independent test set of $2000$ samples.
The three modules are trained using Adam with batch size $48$ and learning rate $10^{-4}$.
Experiments are conducted using PyTorch 2.4.1 on an NVIDIA RTX 4090 GPU.

\textit{Baselines and evaluation protocol:}
For target enumeration, in addition to the traditional cell-averaging constant false alarm rate (CA-CFAR) scheme used in an existing dual-band sensing method~\cite{Liu2025CA}, we further compare DMSNet with advanced signal-processing and learning-based methods, including ADVI-CFAR~\cite{Tian2025-ADVI-CFAR}, CFARNet~\cite{Liang-2025-CFARNet}, Eigenvalue-1D-CNN~\cite{Deng-2026-1D-CNN}, and CSIYOLO-based.
For parameter estimation, five existing dual-band methods are considered and, for brevity, are denoted according to their principal processing mechanisms as Peak-level~\cite{Wei2024CA}, Data-level~\cite{Liu2025CA}, Symbol-level~\cite{Liu2025CA}, Subspace-based~\cite{zhangCA}, and CP-based~\cite{CaiCA}.
The CSIYOLO-based method is also included for comparison.
CSIYOLO-based denotes a dual-band adaptation of CSIYOLO~\cite{zhang2025csiyolo}, where band-specific features are aligned and fused while preserving its original multi-scale detection and parameter-regression mechanisms.
For overall runtime comparison, CA-CFAR is prepended to dual-band methods~\cite{Wei2024CA,CaiCA,zhangCA} without native enumeration, with its latency included.

\subsection{Target Enumeration Performance}\label{se4-b}

Table~\ref{tab:count_performance_0db} compares the target enumeration performance at a representative SNR of $0$ dB using exact count accuracy, MAE, and Macro-F1, where Macro-F1 averages the class-wise harmonic mean of precision and recall~\cite{zhang2025csiyolo}. Compared with the strongest baseline, DMSNet improves Count Acc. from 89.01\% to 91.74\% and Macro-F1 from 90.80\% to 93.07\%, while reducing MAE by 25.8\%.

\begin{table}[t]
    \centering
    \caption{Target enumeration performance at $\mathrm{SNR}=0$ dB.}
    \label{tab:count_performance_0db}
    \small
    \setlength{\tabcolsep}{3pt}
    \renewcommand{\arraystretch}{0.95}
    \begin{tabular}{lccc}
        \toprule
        Method & Count Acc. & Macro-F1 & MAE \\
        \midrule
        Traditional CA-CFAR & 0.7119 & 0.7383 & 0.2999 \\
        ADVI-CFAR           & 0.6546 & 0.6917 & 0.3691 \\
        CFARNet             & 0.7848 & 0.8169 & 0.2309 \\
        Eigenvalue-1D-CNN   & 0.8597 & 0.8831 & 0.1490 \\
        CSIYOLO-based       & 0.8901 & 0.9080 & 0.1123 \\
        Proposed DMSNet     & \textbf{0.9174} & \textbf{0.9307} & \textbf{0.0833} \\
        \bottomrule
    \end{tabular}
\end{table}

\begin{figure}[t]
    \centering
    \includegraphics[width=0.3\textwidth]{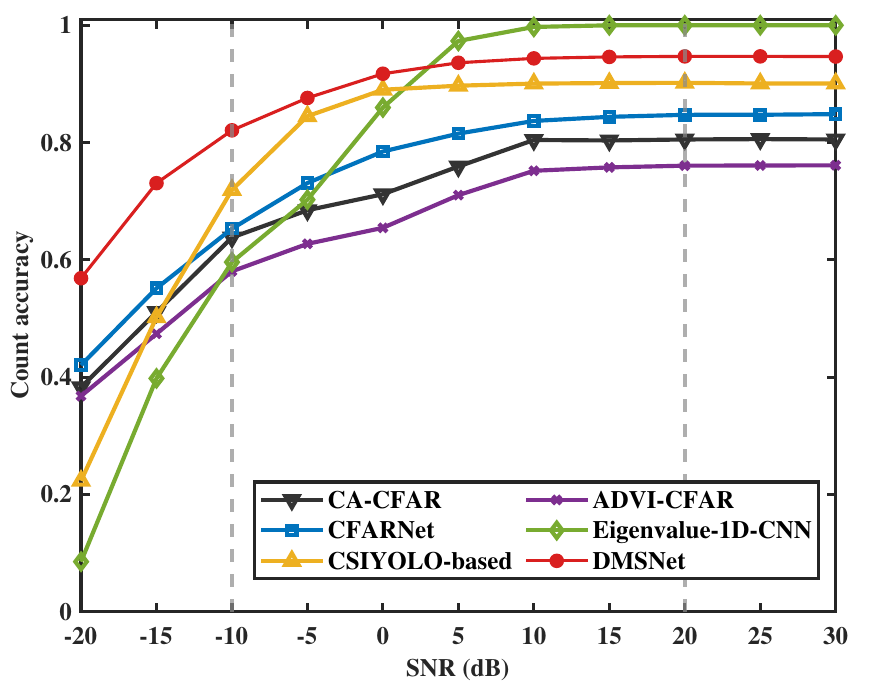}
    \caption{Target count accuracy versus SNR. The vertical dashed lines indicate the training SNR interval.}
    \label{fig:count_snr}
\end{figure}

To further evaluate robustness across different sensing conditions, Fig.~\ref{fig:count_snr} presents the exact count accuracy from $-20$ to $30$ dB. DMSNet achieves the highest accuracy from $-20$ to $0$ dB and degrades gracefully below the training SNR range. 
Its performance also remains stable above $20$ dB, demonstrating good generalization to unseen SNRs.

\subsection{Parameter Estimation Performance}\label{se4-c}

\begin{figure}
	\centering
    \subfigure[Range estimation] {\label{fig:cdf_range}\includegraphics[width=0.22\textwidth]{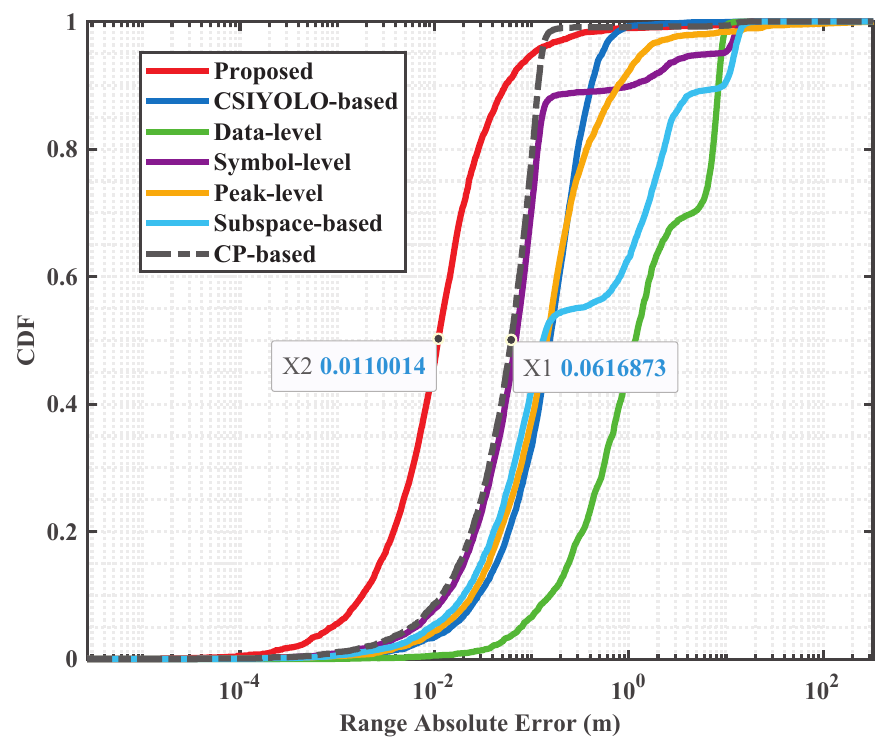}}
    \subfigure[Velocity estimation]{\label{fig:cdf_velocity}\includegraphics[width=0.22\textwidth]{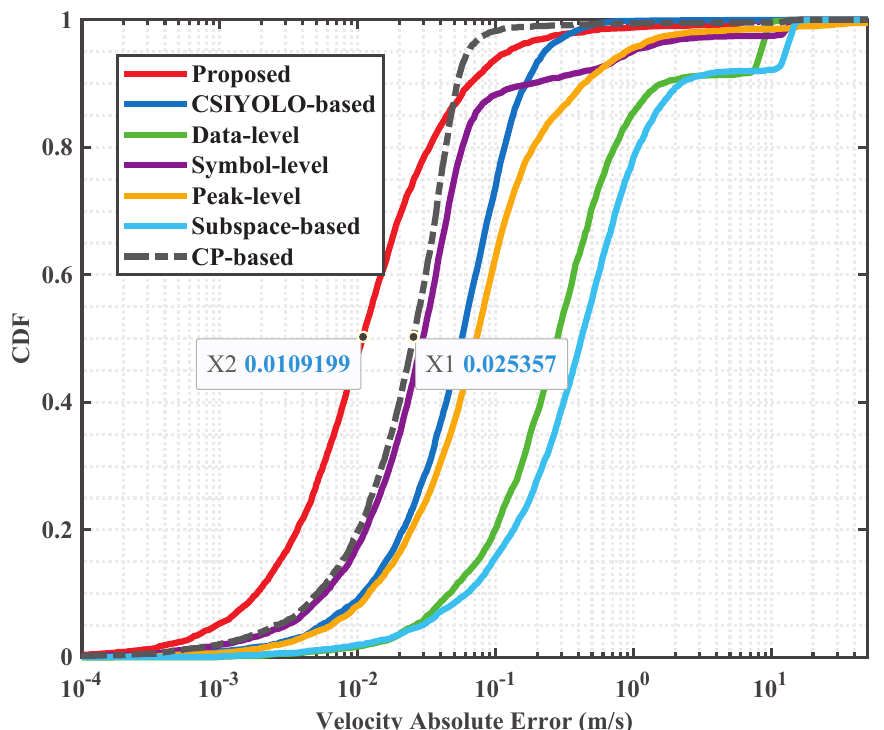}}
    \subfigure[Angle estimation] {\label{fig:cdf_angle}\includegraphics[width=0.24\textwidth]{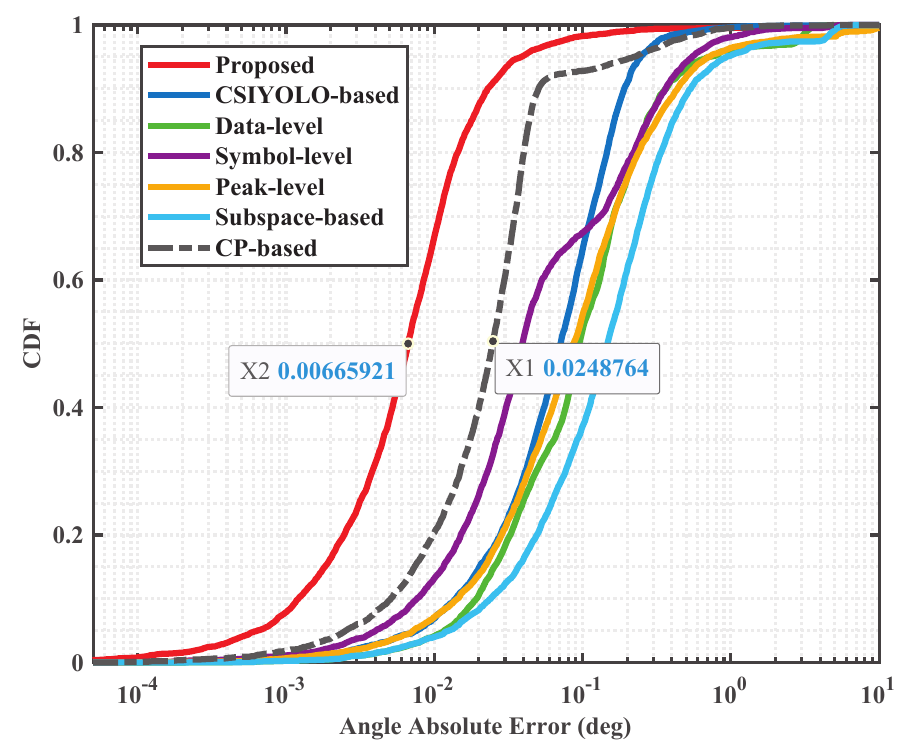}}
	\caption{CDFs of absolute parameter estimation errors.}
	\label{fig:parameter_cdf}
\end{figure}

Fig.~\ref{fig:parameter_cdf} compares the cumulative distribution functions (CDFs) of absolute range, velocity, and angle estimation errors at a representative SNR of $0$ dB.
DMSNet generally shifts the error distributions toward lower values and achieves the best overall performance across the three estimation tasks.
Compared with the best baselines, it reduces the median absolute errors of range, velocity, and angle estimation by 82.2\%, 56.9\%, and 73.2\%, respectively.
This improvement benefits from the coarse-to-fine design, where coarse estimation first identifies the approximate multi-target parameters, and fine estimation subsequently exploits local dual-band information to refine them into continuous-valued estimates.

\subsection{Runtime Comparison}\label{se4-d}

Table~\ref{tab:runtime} reports the end-to-end runtime, where conventional dual-band methods include the preceding CA-CFAR enumeration stage.
All learning-based methods are evaluated with a batch size of one, and only synchronized forward inference is timed.
DMSNet requires only $23.697$ ms, reducing runtime by $68.7\%$ relative to the fastest conventional pipeline, CP-based~\cite{CaiCA}.
CSIYOLO-based is faster owing to its compact end-to-end architecture, whereas DMSNet sequentially executes enumeration, coarse estimation, and fine refinement.
Nevertheless, DMSNet remains substantially faster than conventional pipelines while delivering superior sensing performance.

\begin{table}[t]
\centering
\caption{End-to-End Runtime Comparison.}
\label{tab:runtime}
\small
\setlength{\tabcolsep}{5pt}
\begin{tabular}{lc}
\hline
Method & Runtime (ms) \\
\hline
Data-level~\cite{Liu2025CA}      & 200.535 \\
Symbol-level~\cite{Liu2025CA}    & 188.925 \\
Peak-level~\cite{Wei2024CA}      & 10095.965 \\
Subspace-based~\cite{zhangCA} & 185.025 \\
CP-based~\cite{CaiCA}        & 75.665 \\
CSIYOLO-based~\cite{zhang2025csiyolo}   & \textbf{9.740} \\
Proposed DMSNet           & \underline{23.697} \\
\hline
\end{tabular}
\end{table}

\subsection{Ablation Study}\label{se4-e}
Table~\ref{tab:ablation} compares single-band processing, naive feature concatenation, and the adopted cross-band fusion at a representative SNR of $0$ dB.
Here, P50 and P90 denote the error values below which 50\% and 90\% of the test samples fall, respectively, corresponding to the median and 90th-percentile errors in the CDF.

Compared with high-only, DMSNet improves count accuracy by 7.58 percentage points and reduces the P90 range, velocity, and angle errors by 17.8\%, 18.8\%, and 9.6\%, respectively. 
Compared with naive concatenation, it reduces the velocity P90 by 38.2\% with only 1.85\% more parameters and 0.72\% more floating-point operations (FLOPs), verifying the effectiveness of cross-band fusion.

Furthermore, we evaluate the crucial role of the proposed fine parameter estimation module in Table~\ref{tab:ablation}.
Coarse-only indicates using the coarse estimate directly as the final output. Adding local 3D FFT refinement reduces the P50 range, velocity, and angle errors by 99.5\%, 97.8\%, and 99.3\%, respectively, confirming its importance for continuous parameter recovery.

\begin{table}[!t]
\centering
\caption{Ablation results at a representative SNR of $0$ dB.}
\label{tab:ablation}
\footnotesize
\setlength{\tabcolsep}{2pt}
\renewcommand{\arraystretch}{0.95}
\begin{tabular*}{\columnwidth}{@{\extracolsep{\fill}}lccccc}
\toprule
\begin{tabular}[c]{@{}l@{}}Metric\end{tabular} & \begin{tabular}[c]{@{}c@{}}High-\\only\end{tabular} & \begin{tabular}[c]{@{}c@{}}Low-\\only\end{tabular} & \begin{tabular}[c]{@{}c@{}}Naive\\concat.\end{tabular} & \begin{tabular}[c]{@{}c@{}}Coarse-\\only\end{tabular} & \begin{tabular}[c]{@{}c@{}}DMSNet\end{tabular} \\
\midrule
Count Acc. & 0.8423 & 0.7287 & 0.9092 & \textbf{0.9181} & \textbf{0.9181} \\
$R$-P50 (m) & 0.0109 & 2.1628 & \textbf{0.0106} & 2.1192 & 0.0110 \\
$R$-P90 (m) & 0.0665 & 24.7413 & 0.0551 & 3.9749 & \textbf{0.0546} \\
$V$-P50 (m/s) & 0.0109 & 0.6485 & 0.0121 & 0.4999 & \textbf{0.0109} \\
$V$-P90 (m/s) & 0.0818 & 3.8915 & 0.1075 & 1.0500 & \textbf{0.0664} \\
$A$-P50 ($^\circ$) & 0.0069 & 0.8230 & 0.0069 & 0.9692 & \textbf{0.0067} \\
$A$-P90 ($^\circ$) & 0.0255 & 5.7794 & 0.0241 & 1.7667 & \textbf{0.0230} \\
\midrule
Params (M) & 9.758 & 9.758 & 14.351 & 11.800 & 14.616 \\
FLOPs (G) & 63.97 & 63.97 & 74.70 & 17.75 & 75.24 \\
\bottomrule
\end{tabular*}
\end{table}

\section{Conclusion}\label{se5}
This letter investigates multi-target sensing in heterogeneous dual-band ISAC systems and proposes DMSNet for joint target number and continuous parameter estimation.
DMSNet first estimates the target number and then performs coarse multi-target parameter estimation.
The coarse estimates are refined into continuous range, velocity, and angle parameters.
Simulation results show that DMSNet increases Count Accuracy from 89.01\% to 91.74\% and Macro-F1 from 90.80\% to 93.07\%.
It also reduces target-number MAE by 25.8\%.
For parameter estimation, the median absolute errors of range, velocity, and angle are reduced by 82.2\%, 56.9\%, and 73.2\%, respectively.
In addition, DMSNet reduces runtime by 68.7\% relative to the fastest existing dual-band ISAC sensing method.

% reference
\bibliographystyle{IEEEtran}
\bibliography{reference}

\end{document}